\documentclass[preprint]{elsarticle}  

\usepackage{graphicx}  
\usepackage{dcolumn}   
\usepackage{bm}        
\usepackage{amssymb}   
\usepackage[latin1]{inputenc}
\usepackage[english]{babel}
\usepackage{amsmath}
\usepackage{amssymb}
\usepackage{graphicx}
\usepackage{fancyhdr}
\usepackage{nomencl}
\usepackage{epigraph}
\usepackage{color}

\usepackage[pdfpagelabels,pdftex,final]{hyperref}
\newcommand{\mv}[1]{\mathbf{#1}}

\begin{document}

%
%
%
%
%
\title{Thickness dependence of dynamic and static magnetic properties of pulsed laser deposited La$_{0.7}$Sr$_{0.3}$MnO$_3$  films on SrTiO$_3$(001) }
\author[ntnup]{\AA smund Monsen} \address[ntnu]{Dept. of Physics, NTNU, 7491 Trondheim, Norway}
\author[ntnue]{Jos E. Boschker}\address[ntnue]{Dept. of Electronics and Telecommunications, NTNU, 7491 Trondheim, Norway}
\author[nyu]{Ferran Maci\`a} \address[nyu]{Dept. of Physics, New York University, 4 Washington Place, New York, NY 10003, USA}
\author[ntnup]{Justin J. Wells}
\author[uu]{Per Nordblad} \address[uu]{Dept. of Engineering Sciences, Uppsala University, Box 534, SE-751 21 Uppsala, Sweden}
\author[nyu]{Andrew Kent}
\author[uu]{Roland Mathieu}
\author[ntnue]{Thomas Tybell}
\author[ntnup]{Erik Wahlstr\"om \corref{cor1}}
\cortext[cor1]{Corresponding author: erik.wahlstrom@ntnu.no}

\date{\today}


\begin{abstract}
We present a comprehensive study of the thickness dependence of static and magneto-dynamic magnetic properties of La$_{0.7}$Sr$_{0.3}$MnO$_3$. Epitaxial pulsed laser deposited La$_{0.7}$Sr$_{0.3}$MnO$_3$ / SrTiO$_3$(001) thin films in the range from 3 unit cells (uc) to 40 uc (1.2 - 16 nm)  have been investigated through ferromagnetic resonance spectroscopy (FMR) and SQUID magnetometry at variable temperature. Magnetodynamically, three different thickness, $d$, regimes are identified:  20 uc $\lesssim d$ uc where the system is bulk like, a transition region 8 uc $\le d \lesssim 20$ uc where the FMR line width and position depend on thickness and $d=6$ uc which displays significantly altered magnetodynamic properties, while still displaying bulk magnetization. Magnetization and FMR measurements are consistent with a nonmagnetic volume corresponding to $\sim$ 4 uc. We observe a reduction of Curie temperature ($T_C$) with decreasing thickness, which is coherent with a mean field model description. The reduced ordering temperature also accounts for the thickness dependence of the magnetic anisotropy constants and resonance fields. The damping of the system is strongly thickness dependent, and is for thin films dominated by thickness dependent anisotropies, yielding both a strong 2-magnon scattering close to $T_c$ and a low temperature broadening. For the bulk like samples a large part of the broadening can be linked to spread in magnetic anisotropies attributed to crystal imperfections/domain boundaries of the bulk like film.

\end{abstract}

\maketitle

\section{Introduction }
\label{sec:intro}

Manganites have attracted large attention due to their high spin polarization \cite{park1998, bowen2003}, Curie temperature above room temperature ($T_C$) \cite{tokura1999} and the appearance of colossal magnetoresistance \cite{jin1994}. In particular La$_{0.7}$Sr$_{0.3}$MnO$_3$ (LSMO) has been regarded as a prototypical model system for magnanites, being one of the highest $T_C$ manganites that can be grown. However, it seems as use of the manganites in devices is obstructed by change in properties when interfaced to other materials \cite{sun1996, bowen2003, ogimoto2003, donnell2000}. Accordingly, understanding the surface and interface properties through the study of the thickness dependence magnetic and transport properties have been the focus of several studies. Such studies report on reduced $T_C$ and enlarged coercivity with decreasing thickness, while indicating the existence of inactive and electrically insulating interface layers of thin films \cite{huijben2008, kim2010,Boschker:2012jg} grown on SrTiO$_3$(001), these studies report that very thin layers 3-4 unit cells films (uc) are nonmagnetic, and that they become conductive and magnetic at a thickness of 6-8 uc.

In contrast to the well studied thickness evolution of static magnetic and transport properties, less attention has been paid to magneto dynamic properties, which may provide further information on important length scales for establishment of LSMO bulk properties in thin films. Dynamic properties as investigated through ferromagnetic resonance spectroscopy (FMR) can be used to extract information about the local magnetic environment; FMR line positions and line shapes  can thus provide information on both magnetic state, local anisotropies, and defects \cite{melkov1996,heinrich1993,farle1998}. Studies of the low temperature ferromagnetic phase and magneto dynamics of LSMO using FMR report \emph{e.g.} on phenomena such as spin waves \cite{golos2007, lofland1995b},  magnetic anisotropies \cite{belmeg2010, belmeg2011, infante2008}, and low field adsorption \cite{Golosovsky:2012tx}. These studies also provide evidence for well-defined resonance lines, which is a prerequisite for magnetodynamic devices. However, no concerted effort has been made to map thickness and temperature dependence of the dynamic properties and to understand the physical processes that determine the FMR line-width in LSMO thin films.

In this study  we map out the thickness dependence of dynamic magnetic properties for thin film LSMO on SrTiO$_3$(001) (STO). By linking static and dynamic properties in the thin film region we can explore magnetic anisotropies, damping, coercivity and ordering temperature as a function of film thickness. This gives an opportunity to evaluate the effect of reduced dimensionality and presence of interfaces, on the dynamic magnetic properties of LSMO.

\section{Experimental }
\label{sec:experimental}

\begin{figure}
\includegraphics[width=15 cm]{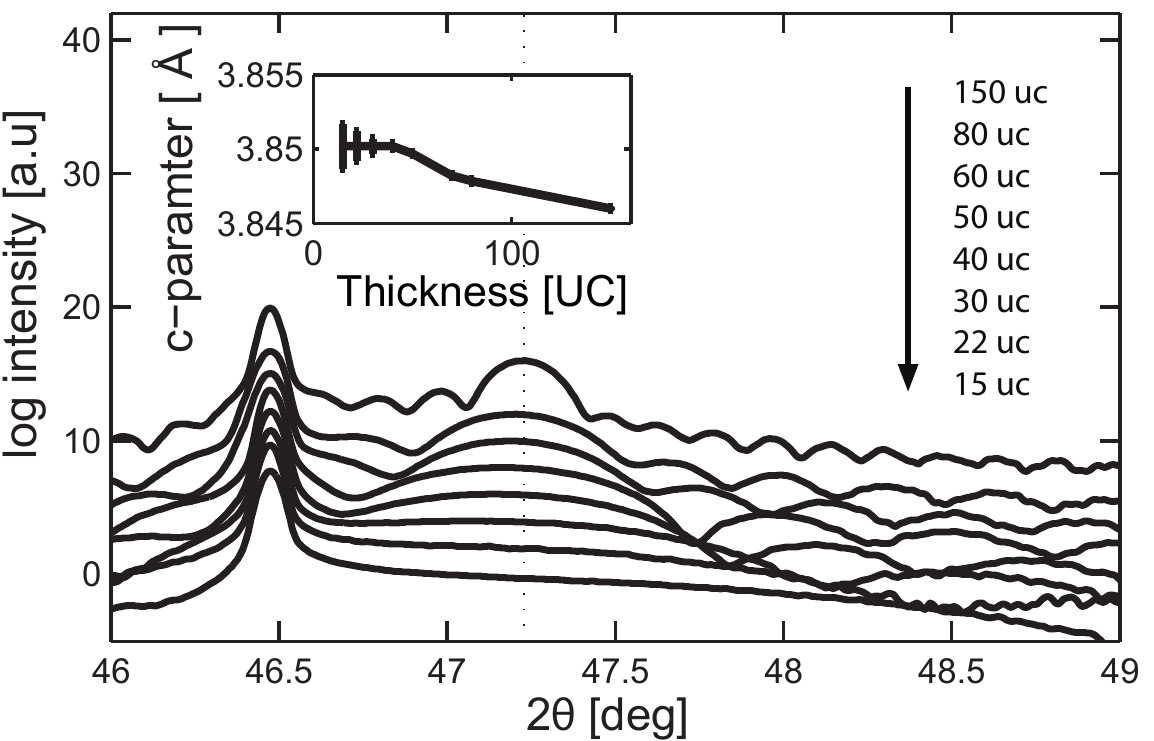}
\caption{\label{fig:XRD} XRD $\theta-2\theta$ scans of the (002) Bragg reflection, revealing an out-of-plane lattice constant of $\sim$ 3.85 \AA\, and thickness fringes indicative of well defined interfaces and constant lattice parameters.}
\end{figure}

A series of LSMO thin films ranging from 3 to 150 uc in thickness were grown epitaxially by pulsed laser deposition (PLD) using a stoichiometric target and a KrF laser ($\lambda=248$ nm) at 1 Hz with a fluency of $\sim$ 2 J/cm$^2$. The as-received TiO$_2$ terminated (001) oriented single crystal SrTiO$_3$ (STO) substrates were prior to deposition annealed at 950$^{\circ}$ C for one hour under flowing oxygen to enhance step quality \cite{lippmaa1999}. The step periodicity for all substrates was similar $\sim$130 nm with no apparent step bunching. During deposition the substrate temperature was set to 680$^{\circ}$ C and held at a distance of 45 mm from the target with a background oxygen pressure of 0.2 mbar. \textit{In situ} reflection high energy electron diffraction (RHEED) was employed to control the thickness and monitor the deposition through observation of mono layer oscillations, consistent with layer by layer growth \cite{rijnders1997, bos2012, bos2013}. A subsequent annealing step was performed \textit{ex situ} under flowing oxygen for two hours at 650$^{\circ}$ C.
Atomic force microscopy (AFM) confirmed the step and terrace morphology of the films after growth, resulting from the vicinal surface of the substrates. The crystalline quality was investigated with four circle X-ray diffraction (XRD). Through fitting to the thickness fringes in the $\theta- 2\theta$ XRD scans around the (002) and (004) Bragg reflections. The films thicknesses determined from the RHEED oscillations were confirmed, and out-of-plane parameters of $\sim$ 3.85 \AA\,  were deduced (Fig. \ref{fig:XRD}).  Reciprocal space maps around the (103) reflection showed a coherently in-plane strained lattice in that direction commensurate with the STO lattice $a=b=3.905$ \AA. The films are grown in growth mode that yields all four possible rhombohedral variants \cite{bos2013}.  The full width half maximum (FWHM) of the rocking curves around the (002) peaks were $\sim  0.03 \,^{\circ}$, comparable to that of the substrates $\sim 0.02\,^{\circ}$. Similar structural parameters are reported for highly optimized LSMO growth \cite{boschker2011}.

Static magnetic measurements were conducted utilizing a Quantum Design SQUID, with temperatures from 10-370 K along the [100] direction. Magnetization versus temperature (M-T) curves were recorded in a field of 50 Oe, after field cooling the samples from 370 K to 10 K in 1000 Oe. Care was taken to eliminate possible magnetic contamination sources \cite{garcia2009}. The diamagnetic contribution from the STO substrates has been corrected for in the analysis of the data.

For dynamic magnetic characterization, samples where cut into $\sim 0.5$x$1$ mm$^2$ pieces using a diamond dicing saw, before loaded with a custom made quartz holder into an x-band (operated at 9.38 GHz) cavity based FMR setup from Bruker (Elexsys II E500) fitted with a 7 kOe electromagnet.  A microwave excitation power of 0.2 mW was employed to avoid saturation and non-linearities, with a 5 Oe modulation of the magnetic field. Angle dependent measurements were performed keeping the $\mv{H}$ field parallel to the film plane using a motorized goniometer. A liquid Nitrogen cooling insert allowed continuous temperature control in the range of 120-380 K. The classic skin depth $\delta =\sqrt{\rho/\mu\omega}$ for LSMO is in the $\mu m$ range \cite{golos2007}, considerably exceeding the film thickness ensuring homogeneous excitation of the ferromagnetic body. The $g$-factor used in the analysis was 2.03 as deduced from a variable frequency flip chip waveguide setup and is close to the values reported for other manganites \cite{causa1998, Ivanshin2000}.

\section{Models}

\begin{figure}
\includegraphics[width=4 cm]{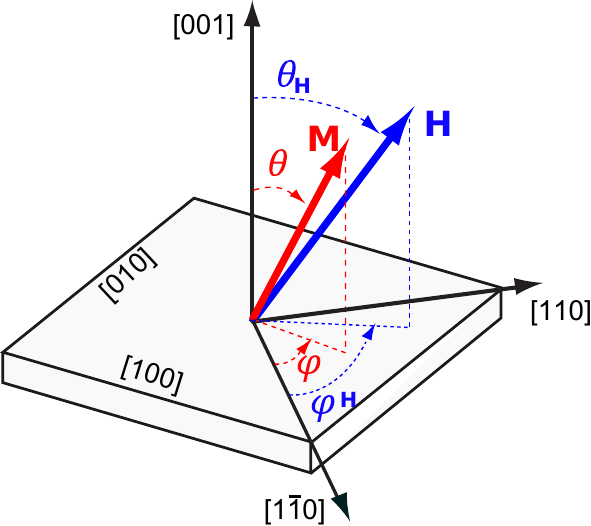}
\caption{\label{fig:coordinate} The coordinate system used in this study. The orientation of the static magnetic field $\mv{H}$ is described by $\theta_H$ and $\phi_H$, and the resulting equilibrium direction of $\mv{M}$ by $\theta$ and $\phi$. }
\end{figure}

We analyzed the data obtained through FMR measurements in the Magnetic-Anisotropy-Energy (MAE) model, which has been successfully applied to metallic ferromagnets \cite{farle1998}, diluted magnetic semiconductors \cite{liu2006}, and oxide thin films \cite{belmeg2010, jalali2005}. In this framework the resonance frequency, $\omega_0$, for uniform precession can be found through the total free magnetic energy of the system \cite{suhl1955, smit1955}. This is the sum of the Zeeman energy ,$E_{zee}=-\mv{MH}$, the demagnetization energy, $E_{demag}= 2\pi M^2 \cos^2(\theta)$, and the magneto crystalline anisotropy energy, where the sample geometry is defined in Fig. \ref{fig:coordinate}.

The magneto crystalline anisotropy energy for a tetragonal system with elongated c axis was used \cite{heinrich1993, farle1998}  valid for both elongation and  contraction. The constants $K_{2\perp}, K_{4\perp}$ and $K_{4\parallel}$ then represent the anisotropy normal and parallel to the (001) plane.
This yields an expression for the total free energy $E_{tot}=E_{zee}+ E_{demag} +E_{mc}$, and the expression for the resonance frequency becomes
\begin{align}
\label{eq:sol}
\begin{split}
\left(\frac{\omega_0}{\gamma}\right)^2= 	&\left[Hcos(\phi-\phi_H) +2 \frac{K_{4\parallel}}{M}cos4\phi          \right] \cdot  \\		
																					&\left[Hcos(\phi-\phi_H) +2 \left(2\pi M	- \frac{K_{2\perp}}{M}\right)+ \right.\\
																					&\left. \qquad \frac{K_{4\parallel}}{2M} (3	+cos4\phi)	     \right]
\end{split}
\end{align}    
for $\theta=\theta_H=90^\circ$, that is when both $\mv{M}$ and $\mv{H}$ are aligned in-plane  \cite{heinrich1993}. Note that in order to separate the uniaxial contribution $K_{2\perp}$ from the demagnetization field, M needs to be independently determined, e.g. by magnetometric methods.

To find the equilibrium magnetization an iterative fitting procedure was applied in the analysis: first, the raw absorption peaks are fitted to Dysonian \cite{feher1954, dyson1955} derivatives describing both the dispersive and absorptive parts of Lorentzian line-shapes, 
\begin{align}
\begin{split}
\label{eq:dysonian}
\frac{dP}{dH}=\frac{d}{dH}	\left(\frac{\Delta H+\alpha(H-H_{res})}{4(H-H_{res})^2 +\Delta H^2 }+\right.\\
		\left.\frac{\Delta H+\alpha(H+H_{res})}{4(H+H_{res})^2 +\Delta H^2 }	\right),
		\end{split}
\end{align}
where $\Delta H$ is the line-width, $\alpha$ the asymmetry factor and $H_{res}$ the resonance field. 
The extracted parameters were stored as function of angle and temperature. From initial trial values for the anisotropy constants, $K_{2\perp} $ and $K_{4\parallel}$, we calculated the equilibrium angles $\phi$ of $\mv{M}$ as a function of field orientation $\phi_H$ through energy minimization with magnetization $M$ as determined from magnetization measurements. Using this we fitted the angle (now modified) to the resolved resonance field, $H_{res}$,  to the solved Smit-Suhl expression (\ref{eq:sol}), yielding the first iteration of the anisotropy constants. This procedure was repeated in an iterative fashion until the value of the anisotropy constants converges. The robustness of this approach was tested for a large number of initial values, all converging to the same result.


\section{Results}

From SQUID magnetometry we deduce an increased coercive field accompanied by a reduction of the Curie temperature with decreasing film thickness  (Fig. \ref{fig:Mag_comb}a and c). For the thinnest films, the magnetic transition temperatures become increasingly less sharp indicating a less homogeneous system. The saturation magnetization extracted from the M-H curves at 10 K is shown in Fig. \ref{fig:Mag_comb}d and displays an apparent decrease with decreasing thickness.

\begin{figure}
\includegraphics[width=12 cm]{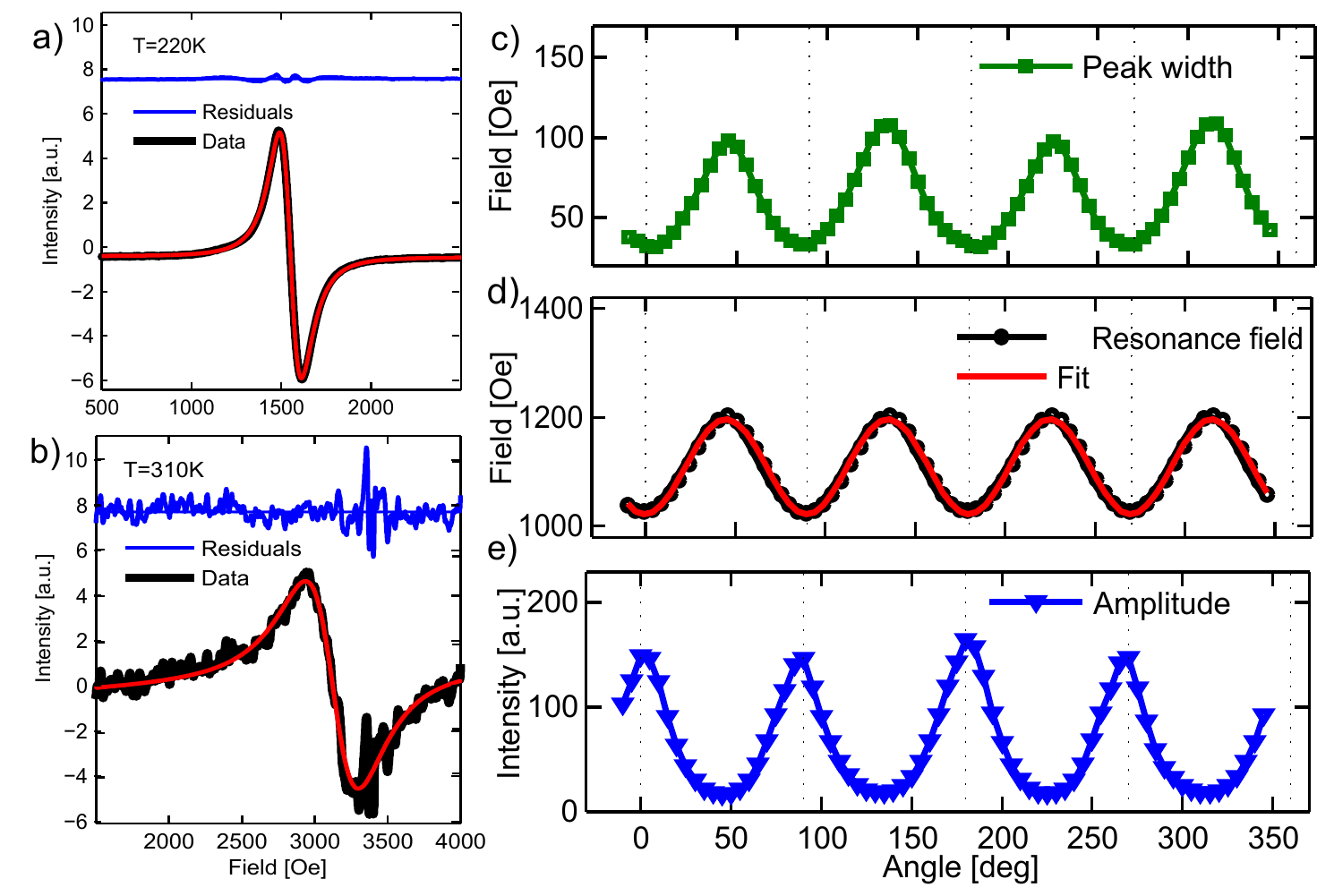}
\caption{\label{fig:rotation} a) Fits of equation \ref{eq:dysonian} to recorded FMR spectra for a 8 uc thin film  at $T<T_C$ and b) $T\sim T_C$ along the [110] direction.  c) The extracted angular dependence of line-width as function of sample angle $\phi$ for the 15 uc film at 200 K, d) resonance field including a fit of equation \ref{eq:sol} and e) amplitude.}
\end{figure}

The thin films of focus here ($d\le 40 $ uc) are well described by a single resonance at all temperatures (Fig. \ref{fig:rotation}a,b). FMR absorption peaks are for all thicknesses well described by the Dysonian derivative (eq. \ref{eq:dysonian}) with no sign of co-existing paramagnetic lines.

\begin{figure}
\includegraphics[width=12 cm]{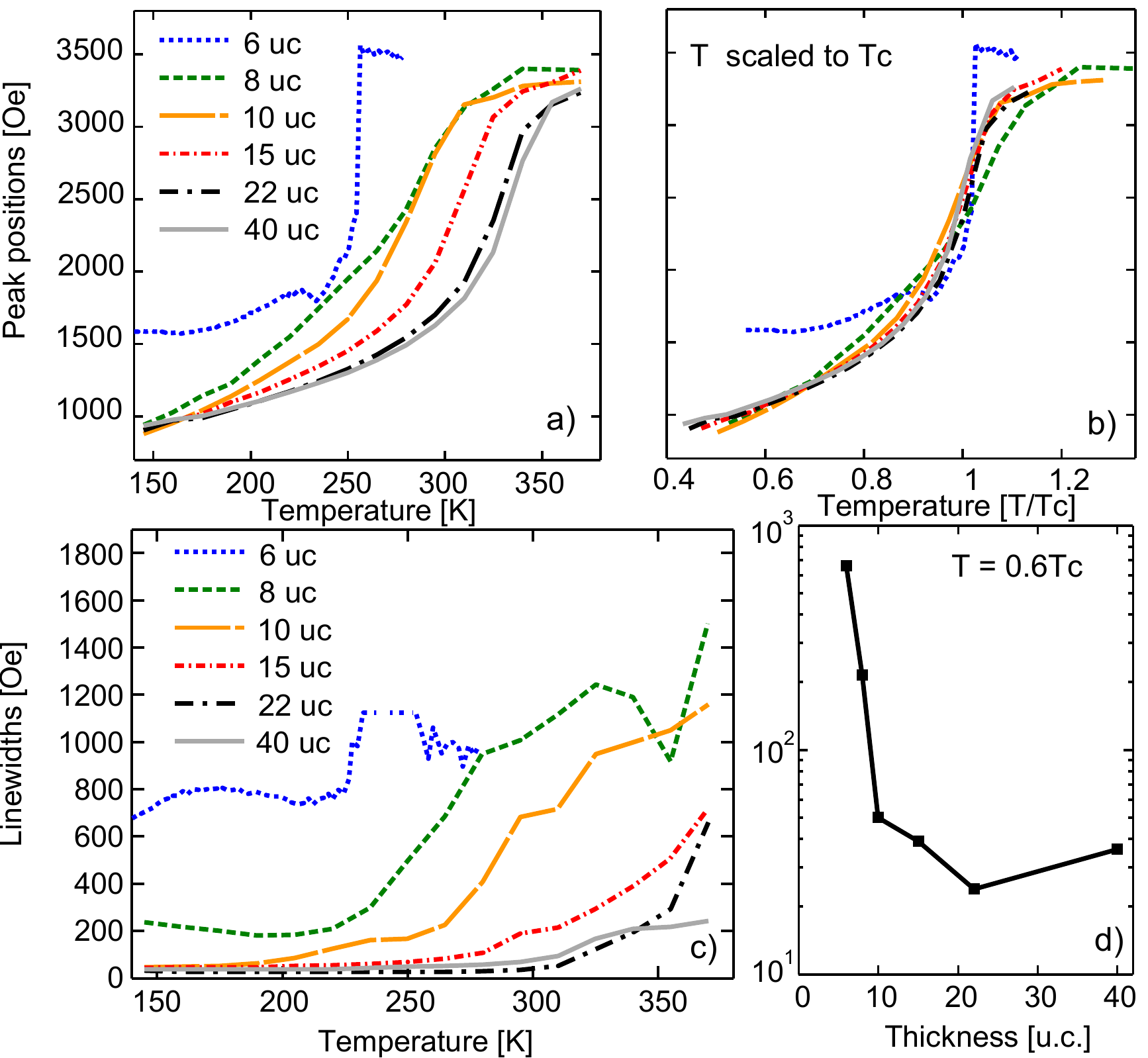}
\caption{\label{fig:pos_width} a) Resonance field dependence on thickness and temperature along easy direction in plane. b) The same resonance data as in a) $T_C$ normalized showing that the shift is mainly due to the reduced $T_C$ for thinner films. In c) the line-width dependence is plotted, and d) the line-width as a function of thickness extracted at 0.6 $T_C$. All data was obtained by fitting of equation \ref{eq:dysonian} to the FMR spectra. }
\end{figure}

The temperature dependence of the resonance fields depends on film thickness (Fig. \ref{fig:pos_width}a), with a more rapid transition to the paramagnetic resonance  ($\sim$ 3400 Oe) with temperature for the thinner films. By renormalizing to $T_C$ (Fig. \ref{fig:pos_width}b), it becomes clear that the data scales with the ordering temperature, apart from  the 6 uc film which clearly separates from the manifold with a field offset of $\sim$ 700 Oe in the low temperature phase and a sharper transition to the paramagnetic phase at the Curie temperature. The absorption peak area of the 6 uc film was also severely reduced compared to the thicker films, contributing only marginally above the background.

Absorption line-widths also displayed a clear thickness dependence (Fig. \ref{fig:pos_width}c). Here, a simple scaling with $T_C$ does not condense the data. However below  $\sim$ 0.8 $T_C$ the line widths become more constant with temperature, and can be represented by Fig. \ref{fig:pos_width}d, where the line-widths extracted at 0.6 $T_C$ are plotted as a function of thickness.

Resonance peaks were also studied under rotation as presented in Fig. \ref{fig:rotation}c-e, displaying clear rotational 4-fold symmetry.
Extracted widths are shown in Fig. \ref{fig:rotation} c) and follow the gradual behavior of the resonance field ( \ref{fig:rotation} d), with maxima in the magnetically hard directions, while the extracted amplitude (ref{fig:rotation} e) is phase shifted by 45 degrees.

\begin{figure}
\includegraphics[width =12 cm]{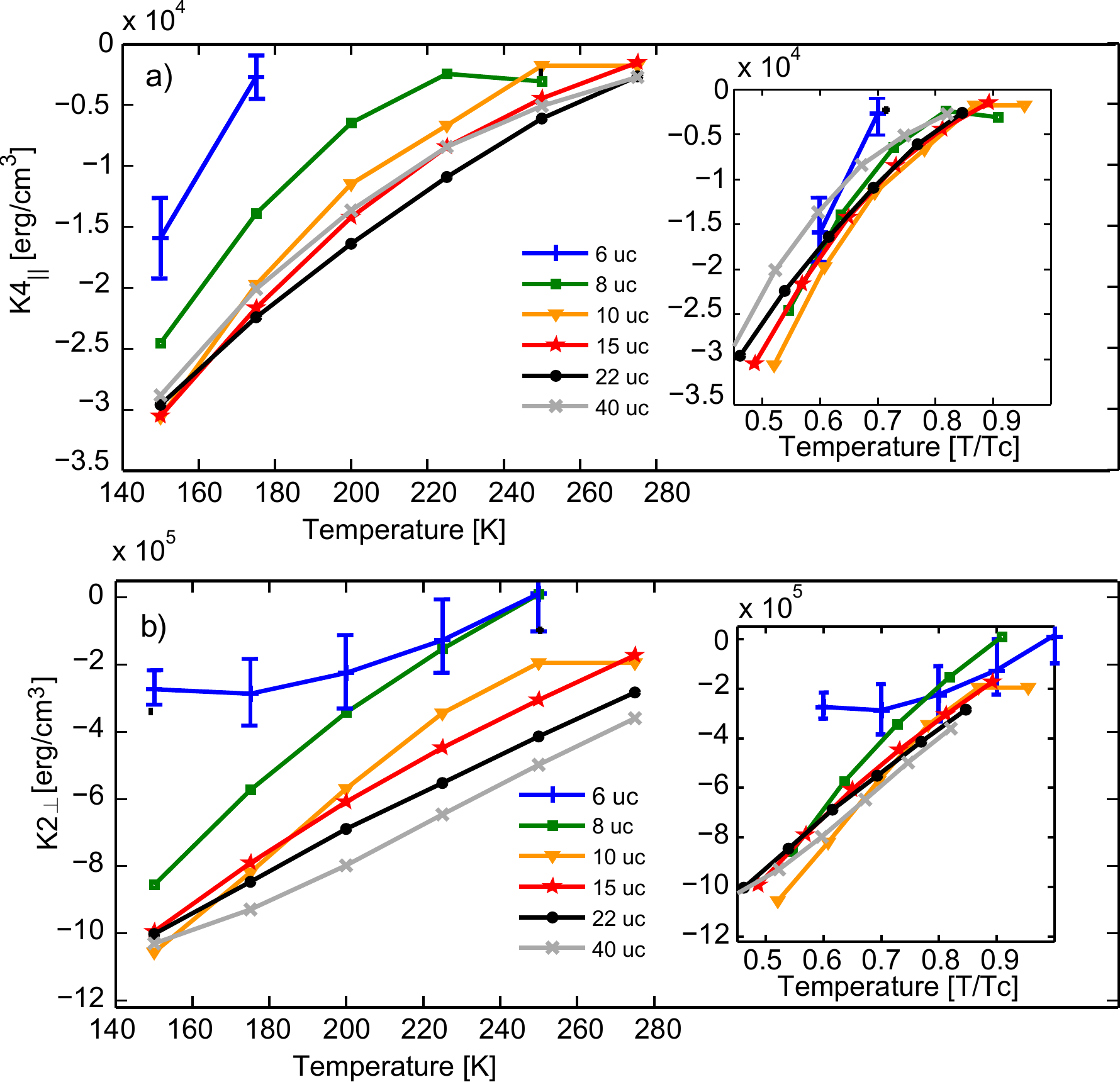}
\caption{\label{fig:anis} The extracted anisotropy constants  a)  K$_{4\parallel}$  and b) K$_{2\perp}$ based on in-plane rotational measurements for all samples, inserts display the same data renormalized by $T_C$ . Specific error bars are drawn for the 6 uc film due to low signal to noise ratio, whereas the uncertainties of the other data points is less than 5\% within a 95\% confidence interval.}
\end{figure}

The in-plane angular dependence of the resonance field indicates magnetic easy axis in the $<$110$>$ directions  and hard axis in the $<$100$>$ in plane directions \ref{fig:rotation} d. The behavior of the resonance field is well modeled by equation \ref{eq:sol}, enabling to extract the anisotropy constants $K_{2\perp}$ and $K_{4\parallel}$ as shown by the fit to data in Fig. \ref{fig:rotation} d. The resulting constants are plotted in Fig. \ref{fig:anis} as function of temperature for all film thicknesses,  including inserts with the same data on a $T$/$T_C$ scale, illustrating that most of the thickness dependent anisotropy is related to the reduced ordering temperature.

The in-plane angular variation of the broadening grows with decreasing temperature as shown by the difference in line width between hard and easy magnetization directions ($\Delta H(\phi)_{max} -\Delta H(\phi)_{min}$), which is summarized as a function of temperature and thickness in Fig. \ref{fig:outwanalys} a.  The out-of-plane (polar angle $\theta$ $\ne$ 90) broadening for the thinnest films could be separated in two distinct  temperature regimes based on the appearance of a cut-off at high angles at temperatures close to $T_C$, as exemplified in Fig.  \ref{fig:outwanalys} b. This high temperature effect dominates at temperatures above $\sim T_C$  and is reduced for thicker films, an example of this is given for the thickness dependence at 0.8 $T_C$  in Fig. \ref{fig:outwanalys} c.

\section{Discussion}

\subsection{Static magnetic properties}

\begin{figure}
\includegraphics[width=6 cm]{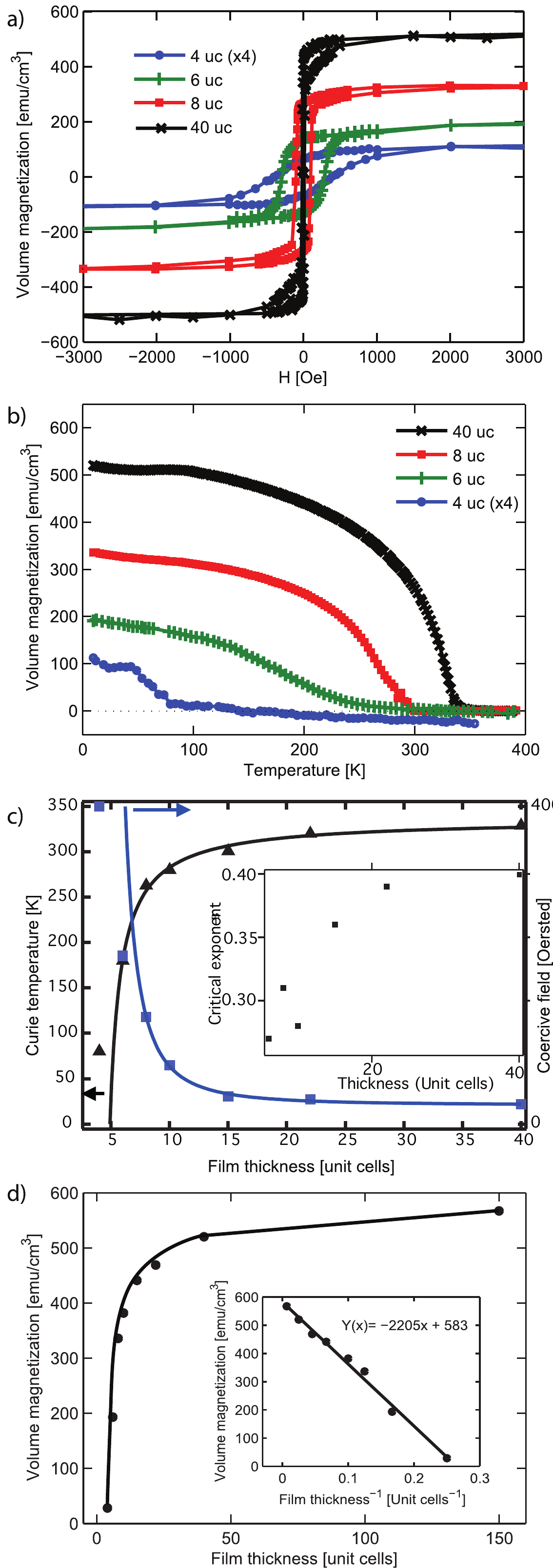}
\caption{\label{fig:Mag_comb} a) M-H plots of the thinnest films recorded at 10 K, displaying thickness dependent coercive fields and saturation magnetizations. Signal from the 4 uc film is scaled by 4 for clarity. b) M-T  data of the same films. c) Summary of  coercive fields and Curie temperatures for all films, with fits as described in the text. (Replotted from \cite{monsen2012}). Insert displays the extracted critical exponents $\beta'$ derived from M-T data. d) Summary of the saturation magnetizations for all thicknesses, including a linear fit to a 1/d plot. }
\end{figure}

The reduced volume magnetization with reduced thickness (Fig. \ref{fig:Mag_comb}d) indicates a magnetically inactive layer. By plotting volume magnetization vs 1/$d$  (insert), we find that a linear relationship with  $\sim$ 4  magnetically inactive uc, and  a constant volume magnetization of $\sim$ 583 emu/cm$^3$ (3.7 $\mu_B$/Mn) describes the system well for all thicknesses. The existence of magnetic dead or altered layers is well documented for LSMO thin films grown on (001) STO substrates \cite{huijben2008, kim2010, Boschker:2012jg, monsen2012}.

Having ascertained the number of uc in the "dead layer", $d_0$, we compare the thickness dependence of the magnetic ordering temperature of the magnetic portion of the films with that described by a simple mean field model:
\begin{equation}
\label{eq:Tc}
1-\frac{T_C(d)}{T_{C_{bulk}}}=c_T(d-d_0)^{-\gamma_T},
\end{equation}
where $c_T$ is a surface dependent constant and the critical exponent $\gamma_T$ should equal unity in a simple cubic lattice \cite{vaz2008}. From this fit  (black full line in Fig. \ref{fig:Mag_comb}c) we obtain $\gamma_T=0.96$ for $4<d$ uc with $C_0=0.9$ and T$_{C_{bulk}}=336 K$.  The change in Curie temperature is thus well described by a mean field model, through the energy change associated with the reduced coordination sites of the magnetic interfaces.

A rough estimate of the dimensionality of the system can be obtained by extracting the thickness dependence of an exponent similar to $\beta$ according to \cite{vaz2008}:
\begin{equation}
\label{eq:crit}
M(T)\propto (1-\frac{T}{T_C})^{\beta' }.
\end{equation}
$T_C$ was estimated by locating the inflection point of the M-T curve and the exponent $\beta'$ was extracted from fits of these curves to eq. \ref{eq:crit}. $\beta'$ is plotted vs. effective film thickness in the insert of Fig. \ref{fig:Mag_comb}c); there is a reduction of $\beta'$ with reduced thickness ($\sim$ 0.40  - 0.25). The observed trend is consistent with a transition from a 3D to a 2D system \cite{vaz2008}, in agreement with the behavior of the Curie temperature. Although we do not expect exact critical exponent values with our method, we note that the results for thicker films correlate well with earlier estimates of  $\beta$ $(\sim$ 0.40  - 0.45) in this material system \cite{ziese2001, kim2002}.
 
Applying a simple power law fit ($H_C(d) \propto (d-d_0)^{\gamma_H},$) to the coercive field data (full blue line, \ref{fig:Mag_comb}c) we observe that it successfully models the behavior of all films thicker than 6 uc with  $\gamma_H=-2$, pointing toward a more complex contribution than a simple surface term for a  purely static surface contribution which would yield $\gamma_H=-1$. The substantial increase of coercive field with decreasing thickness in this region does not stem from a simple surface contribution as shown by the thickness dependence ($\gamma_H$ = -2), suggesting a change in bulk domain structure. Structural pinning-sites \cite{mathieu2000} such as provided by the stepped substrate, polaronic localization, and the gradual transition to a 2D film are all possible candidates for this.

\subsection{FMR line positions and extracted anisotropies}

For films of thickness 8 uc $\le d\le40$ uc, resonance fields and consequently the magnetic anisotropies coincide after renormalization to $T_C$ (Fig. \ref{fig:pos_width}b and inserts in \ref{fig:anis}). In combination with the singular nature of resonance lines prevailing at all temperatures, this points towards a homogeneous magnetic structure with a constant interface contribution, consistent with the observed static properties. The derived magnetic anisotropies are both in magnitude and orientation consistent with previous reports \cite{suzuki1997, belmeg2010, steenbeck2002}.

\subsection{FMR line broadening}

The temperature dependence of the line-widths approaches a constant thickness dependent value at low temperatures, with a substantial increase upon approaching the Curie temperature (Fig. \ref{fig:pos_width}c). Such behavior with temperature is expected when the order in the magnetic phase breaks up,  consistent with data from LSMO single crystals \cite{lofland1997b}, and the inverse magnetization scaling of the Gilbert damping ($\Delta H = \frac{2}{\sqrt{3}}\frac{G \omega}{ \gamma^2 M}$). The thickness dependence of these line-widths is not explained by the varying Curie temperatures as shown in the $T_C$ normalized broadening of Fig. \ref{fig:pos_width}d, and at least one other mechanism must also be contributing.
 
A possible contribution to increased line-widths for thin films are surface/interface imperfections/scatterers which can induce a direct thickness dependent broadening due to local variations in resonance field, or indirectly through two-magnon processes, both which become more dominating as the thickness is reduced. Such variations can be caused both by inhomogeneities in stoichiometry at the surfaces of the film \cite{monsen2012,Boschker:2012jg}, by the uc stepped surface of the STO substrates, as well as more complex interface effects \cite{monsen2012,Boschker:2012jg}.   

\begin{figure}
\includegraphics[width=7 cm]{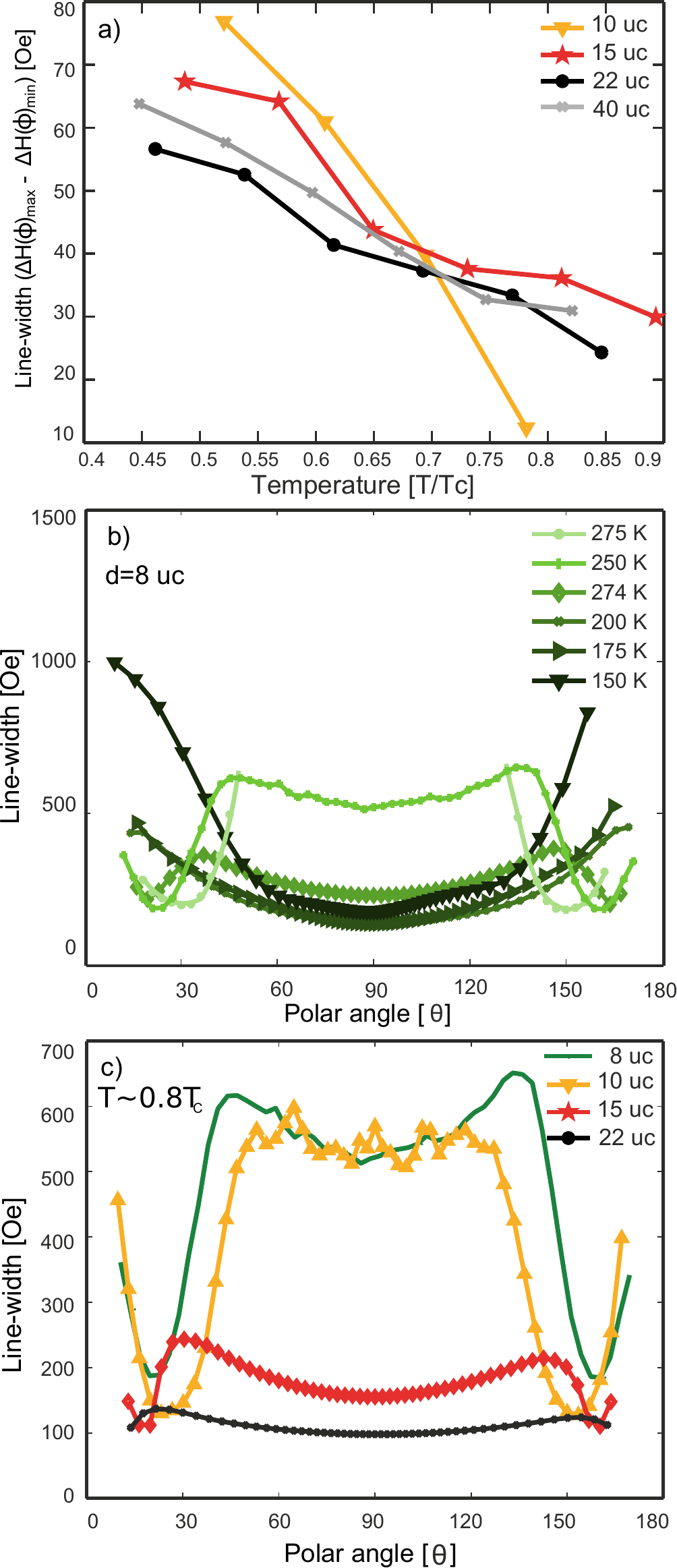}
\caption{\label{fig:outwanalys}a) Temperature dependence of the in-plane line-width due to in-plane magnetic anisotropy ($\Delta H(\phi)_{max} -\Delta H(\phi)_{min}$).
b) Line-widths for the 8 uc film as function of polar angle for a wide range of temperatures.
c) Thickness evolution of the out-of-plane broadening at 0.8 $T_C$, illustrating dependence on thickness. $\theta$= 90 corresponds to in-plane geometry, as defined by figure \ref{fig:coordinate}. }
\end{figure}

At high temperatures and at high out of plane angles we observe a sharp reduction in line-width with polar angle ($\theta <40^\circ$ and $140^\circ <\theta $) of the thinnest films (Fig. \ref{fig:outwanalys}b). This agrees well with two-magnon scattering which depends on the existence of both scatterers of matching wavelength, and spin waves states to scatter into. It has a characteristic cut-off angle $\theta_c$, correlated to the geometric dependence of available states; it is usually set to $\sim 45^{\circ}$, but in practice the cutoff is often positioned higher \cite{Woltersdorf:2004hh}. The two-magnon broadening evidently dominates for thinner films (Fig. \ref{fig:outwanalys}c). We observe an approximate $1/d^2$ dependence in the evolution of the broadening, consistent with the thickness evolution of 2-magnon scattering \cite{Arias:1999tm, Beaujour:2006gz}. In all, it seems to be a dominating mechanism for broadening close to $T_C$ of all thinner samples. 

We also observe an increase in broadening close to the normal alignment of the field (Fig. \ref{fig:outwanalys}b,c ), although we are not able to measure at normal field direction the trend is that this effect is stronger at lower temperatures. We attribute this contribution to the  variation on the local scale of the anisotropy, introducing a broadening that is proportional to the derivative of the resonance field out of plane angle \cite{Zakeri:2007kv}. This term increases at high angles, and as it is thickness dependent we associate it with anisotropy changes due to local surface induced anisotropy variations, which become less important for thick films.

For both rotational axes ($\phi, \theta$) the delta-widths ($\Delta H(\phi)_{max} -\Delta H(\phi)_{min}$ and $\Delta H(\theta)_{max} -\Delta H(\theta)_{min}$), increase with decreased temperatures  but display little thickness dependence, as shown in Fig. \ref{fig:outwanalys} a for the in-plane widths. As the angular values for \textit{max} and \textit{min} correspond to the magnetically hard and easy directions, respectively, we attribute this increase in broadening to the spread in the magnetic anisotropy \cite{Woltersdorf:2004hh, Gurevich:1996wg, farle1998, chappert1986}. This is consistent with the structural domain walls in LSMO thin films running parallel to the magnetically hard $<$100$>$  and $<$010$>$ directions \cite{maurice2003, lebedev2001b,bos2013}. Moreover, the temperature dependence of $\Delta H(\phi)_{max} -\Delta H(\phi)_{min}$  is well correlated with that of the measured magnetic anisotropy constants. 

\subsection{Magnetic thickness regimes in LSMO}

\begin{center}

\begin{table*}[tb]
{\small
\hfill{}
\caption{\label{tab:summary2} A summary of the magnetic properties of LSMO thin films at different thicknesses as deduced from our study. $M$ and $H_C$ extracted at 10 K. }

\begin{tabular}{lcccc}			
\label{table1}Property 	 	& $d=6$ uc 			&	8 uc $\le d \lesssim 20$ uc		&	20 uc $\lesssim d$		$\le$ 40 uc\\
\hline
$M$			&	3.7$\mu_B$/Mn	&	3.7$\mu_B$/Mn		&	3.7$\mu_B$/Mn		\\
$T_C$			& Mean field 			& Mean field 					& Mean field			\\
$H_C$     			&Reduced			&$1/d^2$					&$1/d^2$						\\
K$_{4\parallel}$ / K$_{2\perp}$		&Reduced			&$T_C$ renormalized		&$T_C$ renormalized		 \\

 High t. broadening	&-		& Two-magnon	&  Anisotropy						\\
 Low t. broadening 	&-		& Anisotropy 			& Anisotropy			\\
 \end{tabular}

\hfill{}}
\end{table*}

\end{center}

The overall thickness dependence as summarized  in table \ref{table1} in agrees well to previous reports on magneto static and charge transport.  We observe that the static measurements yields $\sim$3.7$\mu_B$/Mn and a $T_C$ that is well described through a mean field model for all thicknesses, and that all films include magnetically inactive layers corresponding to $\sim$ 4 uc thickness. 

The contrast between dynamic and static magnetic measurements provides new information on the thickness behavior: The very thin film, $d$ = 6 uc, displays significantly altered dynamic properties compared to thicker films yet the volume magnetization indicates that the film still contains a stoichiometric magnetic volume, and it is well modelled by the mean field model describing the reduced $T_C$ (Fig. \ref{fig:Mag_comb}c, d). This discrepancy between the static and the dynamic behavior can be attributed to possible polaron confinement resulting from the reduced dimensionality and/or magnetic anisotropy disorder at the magnetic interfaces. At 6 uc thickness, the magnetic layer should be $\sim$ 2 uc thick, comparable to the polaron size of $\sim$ 3 Mn ions \cite{Chen:2008gs, ziese2001}. Such localization could also be induced by the increased relative contributions of interfaces at this thickness, further driving the trapping. We also note that the abrupt change in dynamic properties from 6 to 8 uc correlates well with the reported ferromagnetic-insulator/ferromagnetic-metal transition occurring at this thickness \cite{kim2010, huijben2008,Boschker:2012jg}, consistent with polaron confinement, which influences charge transport in LSMO \cite{Zener, millis1995}. Given the distinct thickness behaviour of the dynamic properties revealed in this study it would be interesting to directly link it to the charge transport behaviour of the particular thin films. However, it was not possible to address in current study and should be mapped out in a future more dedicated study.

In the transition thickness 8 uc $\le d \lesssim 20$ uc, the thickness dependence of $H_C$, K$_{4\parallel}$, K$_{2\perp}$ and $T_C$ is well described by a mean field model, with a gradual transition into a 3D like system. Dynamically the system is dominated by magnetic surface imperfections, although the anisotropy constants remain constant throughout this transition region. This is concurrent with group B phase identified by Kim \emph{et. al} as a thickness range where a ferromagnetic metallic phase transforms to a paramagnetic insulating phase when crossing the Curie temperature \cite{kim2010}.

\section{Summary}

We have here reported a combined thickness and temperature dependent study of static and dynamic properties of LSMO thin films. A reduced Curie temperature of the thinner films is well described by a mean field model with a constant volume magnetization for all films which accounts for the thickness dependence of $T_C$, $H_{res}$ and magnetic anisotropy constants $K_{2 \perp}$ and $K_{4\parallel}$ that we have extracted from our measurements. 

We distinguish between three different thickness regions. At 6 uc  we have little dynamic coherence, but static ferromagnetic order. In the transition region 8 uc $\le d \lesssim 20$ uc we identify a gradual decrease in line-width with a single resonance peak that indicates an increasingly coherent magnetic phase with a decreasing magnitude of 2-magnon related scattering processes close to $T_C$. At 20 uc $\lesssim d$ the surface contribution to 2-magnon processes is no longer dominating and we observe a observe a bulk like behavior. 

Comparing with previous reports on the thickness dependence of static magnetic properties \cite{huijben2008, kim2010,Boschker:2012jg},  we observe a complementary picture: down to 8 uc, the magnetization and the anisotropies of the films remaining bulk-like indicate a close to stoichiometric composition of the contributing volumes of the sample. The damping seems to be dominated by extrinsic effects: for thin films, locally varying surface induced anisotropy dominates as indicated both by the strong 2-magnon scattering at high temperatures and a high contribution from mosaicity induced broadening at close to normal angles at low temperatures. For thicker samples, a large contribution to the in-plane variation of the broadening is linked to spread in magnetic anisotropies most probably due to crystal imperfections/domain boundaries of the bulk like film.

\section{Acknowledgments}

This work was supported by the Norwegian Research Council (NFR) (project number 182037 and 171332/V30), the Swedish research council (VR) and the Swedish Foundation for International Cooperation in Research and Higher Education (STINT). FM thanks support from EU, MC-IOF 253214.

\addcontentsline{toc}{chapter}{Bibliography} 
\bibliographystyle{apsrev}
\bibliographystyle{unsrt}


\newpage

\newpage

\end{document}